\documentclass{template}
\usepackage{graphicx}
\usepackage{caption}
\usepackage{floatrow}
\usepackage{mathtools}
\usepackage[colorlinks,linkcolor=blue,anchorcolor=blue,citecolor=blue,urlcolor=blue]{hyperref}
\usepackage{marvosym}

\begin{document}
\pagenumbering{arabic}

\title{Asteroid Kamo`oalewa's Journey from the Lunar Giordano Bruno Crater to Earth 1:1 Resonance}

\author{Yifei Jiao$^1$, Bin Cheng$^{1}$\textsuperscript{\Letter}, Yukun Huang$^{1,2}$, Erik Asphaug$^3$, Brett Gladman$^2$, Renu Malhotra$^3$, Patrick Michel$^4$, Yang Yu$^5$ \& Hexi Baoyin$^{1,6}$\textsuperscript{\Letter}}

\maketitle

\begin{affiliations}
    \item Tsinghua University, Beijing, 100084, China
    \item Department of Physics and Astronomy, University of British Columbia, Vancouver, BC, V6T 1Z1, Canada
    \item Lunar and Planetary Laboratory, University of Arizona, Tucson, AZ, 85721, USA
    \item Université Côte d'Azur, Observatoire de la Côte d'Azur, CNRS, Laboratoire Lagrange, Nice, France
    \item Beihang University, Beijing, 100191, China.
    \item Inner Mongolia University of Technology, Inner Mongolia, 010051, China
    \item[\textsuperscript{\Letter}] Email: bincheng@mail.tsinghua.edu.cn; baoyin@tsinghua.edu.cn
\end{affiliations}

\begin{abstract}
Among the nearly 30,000 known near-Earth asteroids (NEAs), only tens of them possess Earth co-orbital characteristics with semi-major axes $\sim$1 au.
In particular, 469219 Kamo`oalewa (2016 HO3), upcoming target of China's Tianwen-2 asteroid sampling mission, exhibits a meta-stable 1:1 mean-motion resonance with Earth.
Intriguingly, recent ground-based observations show that Kamo`oalewa has spectroscopic characteristics similar to space-weathered lunar silicates, hinting at a lunar origin instead of an asteroidal one like the vast majority of NEAs.
Here we use numerical simulations to demonstrate that Kamo`oalewa's physical and orbital properties are compatible with a fragment from a crater larger than 10--20 km formed on the Moon in the last few million years. The impact could have ejected sufficiently large fragments into heliocentric orbits, some of which could be transferred to Earth 1:1 resonance and persist today.
This leads us to suggest the young lunar crater Giordano Bruno (22 km diameter, 1--10 Ma age) as the most likely source, linking a specific asteroid in space to its source crater on the Moon.
The hypothesis will be tested by the Tianwen-2 mission when it returns a sample of Kamo`oalewa. 
And the upcoming NEO Surveyor mission will possibly help us to identify such a lunar-derived NEA population.
\end{abstract}

\section*{Main}
As a close and long-standing companion to our planet, Kamo`oalewa is currently a quasi-satellite (QS) and switches between QS and horseshoe (HS) configurations in a long-term (timescale of Myr) evolution\cite{de2016asteroid,qi2023co}.
{It rotates much faster than most asteroids, period 28.3 (+1.8/-1.3) minutes\cite{sharkey2021lunar}, so must have at least some cohesion}. Its absolute magnitude $H=24.3$\cite{sharkey2021lunar}; for albedos from 0.1 to 0.25 this corresponds to an effective diameter 36--60~m.
{The China Space Station Telescope (CSST)\cite{gong2019cosmology} will be able to make additional observations to help constrain its size and rotation before the arrival of Tianwen-2\cite{zhang2021china}.}
Kamo`oalewa's reflectance spectrum (Fig.~1) provides further clues about its physical properties and compositions: the 1.0-micron absorption band indicates a mineral composition mainly of olivine and/or pyroxene; and the weak band depth and unusual steep red slope is consistent with lunar-style space weathering\cite{sharkey2021lunar,pieters2012distinctive}.
Spectrum curve-matching confirms Kamo`oalewa resembles most closely the lunar soil samples from Apollo 14 (measurement ID: c1lr90) and Luna 24 (c1lr122) missions\cite{relab} (see Extended Data Figs.~1--2).
It also resembles meteorites like Yamato-791197,72 (calmca)\cite{relab,nishiizumi1991exposure} that are fragments from the Moon.
Kamo`oalewa is not compatible with the most common NEA spectra including S-, Q-, and C-types\cite{binzel2019compositional} (as defined by the Bus-DeMeo taxonomy\cite{demeo2009extension}). The less common V-types\cite{migliorini1997vesta}, linked to asteroid Vesta, {have strong absorptions at 1.0 micron and 2 micron, but a modest slope that is inconsistent with Kamo`oalewa.}
The olivine-rich A-types\cite{demeo2019olivine} are rare, and have a much flatter spectrum beyond 1.5-micron.
The striking spectral similarity between Kamo`oalewa and highly-weathered lunar materials, along with its current proximity to the Earth-Moon system, on the other hand, suggests we evaluate a lunar ejecta origin in terms of the cratering event that produced it, and its dynamical route to a co-orbital state. 

If an ancient small body struck the Moon millions of years ago to eject Kamo`oalewa, it would have produced a large number of comparable fragments, among plumes of smaller debris: boulders, rocks, dust, melted and vaporized materials\cite{melosh1989impact,bart2010distributions,horanyi2015permanent}.
The slower ejecta, the bulk of it, would fall back directly, forming the crater rim. Farther-flung ejecta would form bright rays and secondary craters\cite{singer2020lunar,hawke2004origin}.
Ejecta with velocity exceeding the lunar escape velocity, 2.38 km/s, would be efficiently scattered out of the Earth-Moon system by subsequent encounters\cite{gladman1995dynamical}. 
{Most of the ejecta might collide with planets or reside in heliocentric space as NEAs, while a small fraction could reach orbits within Earth's co-orbital zone, indicating that Kamo`oalewa would possibly be amongst such lucky cases\cite{castrocisneros2023orbital}.}

\noindent\textbf{Results}

The size and number of escaping fragments produced by a crater, and the probability of those entering the co-orbital state, imposes constraints that we now evaluate.
We first determine the size of the required crater, and its ejecta properties, by modeling an asteroid impact into lunar target material that is capable of ejecting fragments the size of Kamo`oalewa {at escape velocity}.
We assume an asteroid projectile because the flux of comets is only about $\sim$1\% as great\cite{yeomans2013comparing}.
The impact velocity is expected to be about 18 km/s and the impact angle at 45°, the most probable value\cite{artemieva2008numerical}, which requires 3D simulation.
A further discussion regarding different impact velocities and angles can be found in Supplementary Table~1 and Supplementary Figs.~1--5.
For simulations we use SPH (smoothed particle hydrodynamics) with cohesion and {fracture model assuming an exponential distribution of flaws}\cite{benz1995simulations}.
The Grady-Kipp fragment size distribution is then extended to sub-resolution scales to estimate fragment sizes of damaged SPH particles\cite{grady1980continuum,melosh1992dynamic} (Methods).
The modeling objective is to determine the size-velocity distribution (SVD) of the lunar-escaping fragments, and we look to the Moon for validation. 

The SPH model treats fracture damage as depending on the Weibull parameters $k$ and $m$ {that describe the number of active flaws $n(\epsilon)$ per unit volume for a given elastic strain $\epsilon$, following the power law $n(\epsilon)=k\epsilon^m$.}
Here we calibrate these parameters for lunar materials by trying to match the SVD for Kepler crater, which is 31 km in diameter and provides a good record of its ejecta SVD in the form of a wide distribution of preserved secondary craters\cite{singer2020lunar}. 
We obtain agreement between the upper envelope of the SVD around Kepler, and the results of our SPH simulations, for $k$ from 10$^{30}$ to 10$^{33}$ m$^{-3}$ and $m$ at 9.5 (Fig.~2a).
More generally, we find that the largest expected escaping fragment is in the range 2.5\% to 4.6\% of the projectile diameter. Conversely, an intact escaping fragment 36 m diameter or larger (Kamo`oalewa) requires a crater-forming projectile 0.78--1.44 km or larger, that would produce a crater 10--20 km diameter or larger according to lunar scaling\cite{schmidt1987some,singer2020lunar}.
For a Kepler-sized crater, we would expect about 1250 escaping fragments ($>$36 m) for $k$ = 10$^{30}$ m$^{-3}$, and about 300 for $k$ = 10$^{33}$ m$^{-3}$ (Fig.~2b), with the majority originating from the shallow surface via the process of spallation caused by interaction of stress and rarefaction waves near the free surface\cite{melosh1984impact}.
The median rotational speed of such fragments is close to that of Kamo`oalewa (Extended Data Fig.~3), providing a {possible direct} explanation for its observed super-fast rotation state, {although the fast rotation could also be a recent result by subsequent evolutionary processes such as the Yarkovsky-O'Keefe-Radzievskii-Paddack (YORP) effect, collisions, or gravitational torques\cite{vokrouhlicky2002yorp}}. 

There are tens of thousands of craters larger than 10 km on the Moon\cite{robbins2019new,barlow2017status}, but most of them date to several Gyr\cite{neukum2001cratering}, far older than Kamo`oalewa could plausibly be. Asteroids this small can be subject to disruptive collisions with each other\cite{bottke1994collisional}, and they have short dynamical lifetimes in near-Earth space where they are subject to planetary scatterings and Yarkovsky forces. The median lifetime of NEAs is about 10 Myr, and the longest about 100 Myr, until they collide with a planet or the Sun, or are ejected from the solar system\cite{gladman2000near,huang2021four}.
This would imply that any lunar ejecta presently in near-Earth space would likely derive from a lunar crater that formed within the past 10--100 Myr.

Craters that young, and of the required 10--20 km diameter, are rare over the surface of the Moon, {with only several tens of candidates\cite{mazrouei2019earth} (as listed in Supplementary Table~2), e.g., the widely known lunar craters Giordano Bruno (GB, 22 km, 1--10 Ma\cite{morota2009formation}) and Tycho (85 km, 109 $\pm$ 4 Ma\cite{drozd1977cosmic}).}
Since GB is the youngest of these by a significant fraction, with abundant melt, steep walls, and prominent rays (see Extended Data Fig.~4), we now focus on it as the potential source.
The GB crater presents a pyroxene composition within its inner crater wall and around the rim\cite{bhattacharya2017enhanced}, similar to the mineralogy of Kamo`oalewa.
This is further supported by spectral similarities between Kamo`oalewa and the Luna 24 sample (Fig.~1); the latter has been suggested to contain material from the ejecta from GB\cite{basilevsky2012age}.
While the Kepler crater's estimated formation age of 625--950 Ma\cite{konig1977recent} rules it out as a likely source of Kamo`oalewa, we can directly apply its detailed SPH results to evaluate ejecta characteristics from other lunar craters, using crater scaling (Methods).
We estimate that approximately 100--400 Kamo`oalewa-sized fragments ($>$36~m) would have escaped during the formation of 22-km GB, when the Moon was struck by a 1.66-km asteroid.

Then, can impact ejecta from the GB crater be dynamically transferred into Earth co-orbital space?
It has been reported that there exists a dynamical barrier around $a=$ 1 au for lunar launched particles to enter Earth's co-orbital space; however, a narrow range of launch conditions does allow QS-HS outcomes similar to Kamo`oalewa's orbit\cite{castrocisneros2023orbital}.
Here we perform N-body simulations investigating the orbital evolution of escaping GB ejecta, initialized with various launch velocities, azimuths, and lunar phases, and terminated until 10 Myr or by other events (see Methods and Extended Data Fig.~5).
Each simulation includes 300 particles launched from the coordinates of GB, with velocities and azimuths set according to our impact simulation results, and with a random lunar phase.
{All particles are treated as massless points, even if they are larger than 36 m in size.}
We use a total of 200 sets for 10 Myr simulations to statistically characterize the orbital evolution of the GB ejecta.
As shown in Supplementary Table~3, the majority (about 84\%) of escaping fragments survive and migrate into heliocentric orbits within the first 100 years, while only $\sim$18\% of them survive to 10 Myr, and 0.7\% to 100 Myr (Fig.~3c).
Only a small proportion of the surviving particles become Earth co-orbitals (e.g., Fig.~3a).
{The statistics of Earth co-orbital fractions span several orders of magnitude due to the chaotic dynamics, which is of around 0.0001\% to 1\% during 1--10 Myr (Fig.~3b).}
On average, GB formation is likely to produce around 0.3-1 Earth co-orbitals (0.1\% to 0.3\% of the 300). If the GB event is relatively recent and circumstances are favorable, it could generate as many as 3 Kamo`oalewa-like objects (1\% of the 300).
Note that the Earth co-orbital fraction does not follow a simple power law over time, due to the steep decrease of survival fraction after 10 Myr (Fig.~3c).
Given the survival fraction of 40--18\% within 1--10 Myr, there will be about 50--120 of the 300 ejected fragments ($>$36 m) surviving in near-Earth space and remaining unidentified or undiscovered, which is consistent to the current discovery efficiency of NEAs at this size ($\sim$1\%)\cite{harris2015asteroid}.
Taking into account the large number of smaller escaping fragments (more than 10$^4$ objects $>$10 m as shown in Fig.~2b), we predict thousands of ten-meter-sized GB ejecta are still preserved in space as NEAs.
Although constituting only a small fraction of the asteroid population, future search by NEO Surveyor\cite{mainzer2021near} for this enigmatic class of asteroids with a 1.0-micron absorption band but a steep red slope would provide evidence for the existence of a lunar-ejecta family.

On the other hand, most ejecta end up colliding with the Earth (Fig.~3d); this would generate a roughly ten-fold spike in lunar meteorites delivered to Earth for the first $\sim$1 Myr after GB formation (Extended Data Fig.~6).
This seems to be in tension with the vast majority of lunar meteorites possessing cosmic-ray exposure ages much shorter than the 1--10 Myr age of the Giordano Bruno crater\cite{fritz2012impact}.
However, the spike declines rapidly with time and would be unobservable now due to the short terrestrial preservation of lunar meteorites of only a few hundred thousand years.
Presently, GB ejecta contributes to about 10\% of the background flux of lunar meteorites, implying that our current collection of lunar meteorites likely contains only a handful of GB ejecta, e.g., the possible GB meteorites Yamato-82192/82193/86032\cite{fritz2012impact}.

Combining our SPH and dynamical N-body simulations, we can estimate the evolutionary path of the escaping fragments from any potential source crater on the Moon, as well as their survival and Earth-co-orbital fractions over time (Methods).
We use the standard deviation of the co-orbital fraction distribution as shown in Fig.~3b to describe the uncertainty in the orbital evolution of lunar ejecta.
This allows us to constrain the source crater's size and age to obtain at least one Earth co-orbital object at the present day (Fig.~4).
It is clear that the largest, youngest craters (upper left corner in Fig.~4) are more probable sources, as they produce more escaping fragments that still stay in space or the Earth co-orbital region.
And indeed GB is the only possible source crater satisfying the criterion.

\noindent\textbf{Discussion}

Compared to lunar ejecta, main-asteroid-belt-originated bodies dominate in the NEA population {and also in the Earth co-orbitals}.
{Among the total 250,000 NEAs with $H<$ 24.3, it is anticipated that around one hundred Earth co-orbitals trace their origins to the main asteroid belt\cite{morais2002population}, specifically the inner belt\cite{granvik2018debiased} which is mainly composed of S- and C-type asteroids\cite{demeo2014solar}. However, the spectral inconsistency of Kamo`oalewa may rule it out from such objects, suggesting instead a piece of lunar debris.}

We have explored the processes for impact-induced lunar fragments migrating into Earth co-orbital space, and presented support for Kamo`oalewa's possible origin from the formation of Giordano Bruno crater a few million years ago.
This would directly link a specific asteroid in space to its source crater on the Moon, and suggests the existence of more small asteroids composed of lunar materials yet to be discovered in near-Earth space.
If Kamo`oalewa is indeed lunar ejecta, it would be a precious sample to help us understand the impact-ejection process, and space weathering on the Moon during the past several million years, and in the near-Earth dynamical environment.
The large fragment sizes predicted by our SPH simulations correspond to relatively low shock levels at a distance of kilometers from the impact site, so Kamo`oalewa or its predecessor may remain cohesive when ejected.
Over the course of several million years, thermal fatigue and meteoroid impacts could have further modified its monolithic structure, and produced fractures and grains on its surface\cite{delbo2022alignment,zhang2021shapes}.
Considering its anomalously fast rotation, we predict that Kamo`oalewa could be a young monolithic-type asteroid with possible surface fractures and only shallow regolith\cite{cambioni2021fine}, rather than a classical rubble-pile asteroid like Itokawa and Bennu\cite{rozitis2014cohesive,lauretta2019unexpected}.
Asteroids tens-of-meters in size have never been explored by space missions, and thus are among the least understood small bodies, even though they represent the most frequent NEA hazard and could be the most accessible space resources.
China's Tianwen-2 mission, an in-situ and sample-return mission to Kamo`oalewa\cite{zhang2021china}, 
and the NEO Surveyor mission\cite{mainzer2021near}, that will provide the most detailed survey to data of NEA populations,
will test our hypothesis and provide greater insights into the formation and evolution of our neighboring celestial small bodies.

\begin{methods}
\subsection{Spectral comparisons.}
We compare the reflectance spectra of Kamo`oalewa\cite{sharkey2021lunar}, lunar materials\cite{relab}, and typical asteroid taxonomies\cite{demeo2009extension} with regard to their slopes and absorption features.
The online Bus-DeMeo taxonomy tool\cite{demeo2009extension} is used to smooth, sample and normalize the input spectrum spanning the entire wavelength range 0.45 to 2.45 microns, and to calculate the mean slopes (compared with 24 classes in Bus-DeMeo system in Extended Data Fig.~1).
The smoothed spectra are then represented as latent scores based on the Mixture of Common Factor Analysers (MCFA) using the \textit{classy} package\cite{mahlke2022asteroid}.
Extended Data Fig.~2 shows the latent distributions of Kamo`oalewa and lunar materials, whose spectra have been plotted in Fig.~1, in comparison with 2983 observations of A-, S-, Q-, C-, and other asteroid types from public data\cite{mahlke2022asteroid}.
The MCFA results indicate that Kamo`oalewa's spectrum is closely aligned with lunar materials but refute any spectral connection with these common near-Earth asteroid types.

\subsection{Lunar impact simulations.}
In this work, all SPH simulations are performed using our shock physics code \textit{SPHSOL}\cite{jiao2024sph}, which is specifically designed for impact simulations in planetary science and has been verified by modelling laboratory impact experiments of shooting nylon projectiles into basalt or pumice targets\cite{jiao2024sph}.
The {nominal} scenario is a 2.44-km-diameter spherical asteroid impactor striking the lunar crust at 18 km/s and 45$^\circ$, which will produce a Kepler-sized final crater.
Varying lunar impact conditions leads to ejection results of the same order of magnitude, and thus has minor effects on the required crater size and Earth co-orbital outcomes, as discussed in Supplementary Table~1 and Supplementary Figs.~1--5.
We use a half-sphere target domain with a diameter of 30 km, which should be large enough to capture the behavior of high-speed fragments at the early stage of crater formation (with the transient crater of about 10 km).
With a spatial resolution of 100 m (validated as a reasonable resolution to balance the modeling accuracy and computational cost), the impactor and target are modeled using about 3.6$\times$10$^3$ and 3.5$\times$10$^6$ SPH particles, respectively.
The symmetric boundary condition using ghost particles\cite{bui2008lagrangian} is applied on the symmetry plane defined by the impact velocity vector and target surface normal, to reduce the computational cost.
We assume homogeneous basaltic materials for both the impactor and target, implemented with the Tillotson equation of state\cite{tillotson1962metallic}, the pressure-dependent yield criterion for plastic deformations\cite{collins2004modeling,jutzi2015sph}, and the Grady Kipp model for fragmentation\cite{grady1980continuum,melosh1992dynamic,benz1995simulations} (parameters shown in Supplementary Table~4).
Here we model the target as a continuum with Grady-Kipp flaws to approximate the clast sizes, which is validated by reproducing Kepler's secondary craters.
Also note that the concerned GB crater is melt rich with steep walls, and the rocks are not noticeably layered. These might justify considering a simple Grady-Kipp target in this work.
For each SPH particle $i$ with the peak fragment size $L_m$, the cumulative fragments number is computed as\cite{melosh1992dynamic}
\begin{equation}\label{eq_Fcumi}
\setlength{\jot}{12pt}
\begin{split}
    F_{\mathrm{cum}}^i(L) & = \frac{(m+6)(m+5)(m+4)}{120}\frac{V_{\mathrm{cell}}}{L_{\max}^3}\left(1-\frac{L}{L_{\max}}\right)^m\left[1+m\left(\frac{L}{L_{\max}}\right)\right.\\
    &\left.+\frac{m(m+1)}{2}\left(\frac{L}{L_{\max}}\right)^2+\frac{m(m+1)(m+2)}{6}\left(\frac{L}{L_{\max}}\right)^3\right]
\end{split}
\end{equation}
where $L_{\max}=\frac{m+2}{3}L_m$ and $V_{\mathrm{cell}}$ is the cell volume of the particle. The total fragment size distribution is then the summation of individual size distributions as $\sum F_{\mathrm{cum}}^i(L)$. 
The largest probable fragment size is determined as
\begin{equation}\label{eq_Fcum}
    \sum F_{\mathrm{cum}}^i(L_{\mathrm{largest}}) \equiv 1
\end{equation}

Also, we calculate the vorticity magnitudes of SPH particles from the rotation tensor, which is the anti-symmetric part of the velocity gradient, to represent the rotation distribution of sub-resolution fragments.

\subsection{Lunar crater scaling law.}
The crater scaling law has been used to relate a lunar crater to its possible impactor\cite{schmidt1987some,singer2020lunar},
\begin{equation}\label{eq_scaling1}
    D_{\mathrm{imp}} = 1.004 D_{\mathrm{tr}}^{1.275} \left [ \frac{g}{\left ( v \sin\theta  \right )^2 }  \right ]^{0.275} 
\end{equation}
where $D_{\mathrm{imp}}$ is the impactor diameter and $D_{\mathrm{tr}}$ is the transient crater diameter (in km). $g$ is lunar gravitational acceleration, $v$ is the impact velocity, and $\theta$ is the impact angle (measured from the ground plane). The gravity collapse of a transient cavity will make a larger final crater $D_{\mathrm{final}}$, which is estimated from\cite{singer2020lunar}
\begin{equation}\label{eq_scaling2}
    D_{\mathrm{tr}}=1.15D_{\mathrm{final}}^{0.885}
\end{equation}

Considering the impact velocity at 18 km/s and 45$^\circ$, we estimate that a 2.44 km impactor will produce a 31 km final crater (Kepler sized), and a 1.66 km impactor for a 22 km crater (GB sized).
The escaping fragments number of other sized craters can be estimated with reference to the Kepler crater via the assumed linear relation between the escaping ejecta and the impactor\cite{artemieva2008numerical} as
\begin{equation}\label{eq_escape_imp}
    N_{\mathrm{Launch}} \propto D_{\mathrm{imp}}^{3}
\end{equation}
which has been validated by the supplemental impact simulation of a GB-sized crater (see Supplementary Fig.~6).

\subsection{N-body simulations.}
The open-source software package \textit{REBOUND}\cite{rein2012rebound} is used for our N-body simulations, with its hybrid integrator \textit{MERCURIUS}\cite{rein2019hybrid}.
{The \textit{MERCURIUS} uses an implementation of the symplectic second-order Wisdom-Holman integrator WHFast\cite{rein2015whfast} for long-term integrations of planetary systems and switches over smoothly to the higher-order integrator IAS15\cite{rein2015ias15} during close encounters, which integrates with a smaller, adaptive time step. Therefore, the integrator is capable to simulate the complex dynamical evolution of lunar ejecta over long-time periods, e.g., 10--100 Myr, with optimizing both speed and accuracy.}
We have estimated 100--400 Kamo`oalewa-sized escaping fragments from a GB-scale impact event.
To investigate the subsequent orbital evolution of these objects, we perform {200} sets of N-body simulations, with each set including 300 massless particles (a reasonable value between 100 and 400) launched from the GB crater (35.9° N, 102.8° E).
As illustrated in Extended Data Fig.~5, we initialize each set at a random lunar phase, which indicates the relative position of the Sun-Earth-Moon system, and launched 300 particles along a 45° cone at random azimuths.
The initial velocities of these particles adhere to a given velocity distribution (a power law distribution ranging from 2.38 to 6.0 km/s, and with the power of $-$4.0) as predicted by our SPH simulations in Supplementary Fig.~7.
As such, each of the 200 simulation sets represents a numerical reenactment of the GB impact event.
In the first few decades, the motion of these particles is dominated by the gravity of the Moon and Earth, while also being strongly perturbed by the Sun\cite{gladman1995dynamical}.
We model this geocentric stage as a four-body problem consisting of the Sun, Earth, Moon, and ejected particles, and integrate the system for 100 yr, by which time the vast majority of particles have escaped from the Earth's Hill sphere.
For the subsequent heliocentric stage, we include all the eight planets, the Sun, and test particles as a full N-body model and integrate the system for 10 Myr, where the Earth-Moon system is treated as a single body at their barycenter\cite{gladman1995dynamical}.
We export the all particles' orbital elements every 100 yr for the heliocentric stage.
The JPL planetary and lunar ephemerides are used to construct initial conditions for all the massive bodies\cite{park2021jpl}.
Note that we are focusing on lunar fragments in tens of meters sized, which may have a moderate thermal conductivity like lunar sub-surface or meteorites, on the order of 1 W/(K$\cdot$m)\cite{delbo2015asteroid}.
Thus the Yarkovsky-induced semi-major axis drift rates are on the order of $10^{-3}$ au/Myr for such objects\cite{fenucci2021role}, which is negligible and not included in our simulations.
{For ten-meter-sized or larger rocks while in space, solar radiation pressure forces are negligible over lifetimes of $<$100 Myr\cite{burns1979radiation}.}
Particles are removed from the integration if they collide with planets or the Moon, cross the Jupiter's orbit, or become Sun-grazers.

The 10-Myr N-body simulations are sufficient to model GB ejecta's dynamical evolution, but provide limited information for older crater cases (e.g., the 109-Ma Tycho crater).
However, simulating the ejecta evolution of all possible source craters is quite computationally expensive, due to their diverse locations and extended ages.
Consequently, we conduct {20} sets of additional simulations (each of 300 particles) for a total of 100 Myr, following the same conditions as described before.
In summary, our N-body simulations consist of {60,000} particles for 10 Myr and {6,000} particles for 100 Myr, both launched from the GB crater.
The results are presented in Supplementary Table~3.
As a comparison, previous works on lunar ejecta evolution found that roughly a quarter to a half of lunar ejecta (from random site on lunar surface) will collide with the Earth within 10 Myr\cite{gladman1995dynamical}.
The fraction of Earth collisions in our simulations is slightly larger, about 60\%. 
This observation is in line with previous indications that lunar ejecta originating from the trailing side are more likely to produce Earth-like orbits and also Earth collisions\cite{castrocisneros2023orbital}.

To recognize particles' orbital state, we process the raw data of the semi-major axis using a low-pass filter (e.g., the red dashed line in Fig.~3a).
Then we limit the filtered semi-major axis from 0.998 to 1.002 au to find Earth co-orbitals like Kamo`oalewa.
And the mean longitude relative to Earth $\Delta\lambda$ provides information about co-orbital configurations: QS ($\Delta\lambda$ librates around 0$^\circ$), HS ($\Delta\lambda$ librates around 180$^\circ$), or TP ($\Delta\lambda$ librates around $\pm$60$^\circ$).
Since the particles' orbits are subject to strongly chaotic dynamics, the real-time number of Earth co-orbiters is not convincing.
Here we use the average value over a time interval $\Delta T$ to evaluate the proportion of the Earth co-orbitals to the launched particles,
\begin{equation}\label{eq_define_R2L}
    R_{\mathrm{1:1\ E\ to\ Launch}}(t) = \frac{\sum _{t-\frac{\Delta T}{2}}^{t+\frac{\Delta T}{2}} \frac{\Delta t_i}{\Delta T} N_{\mathrm{1:1\ E}}^{t_i}} {N_{\mathrm{Launch}}}
\end{equation}
where $N_{\mathrm{Launch}}=300$ in the N-body simulations, and $\Delta t_i$ is the output time step of 100 yr. For each output step, the Earth co-orbitals number $N_{\mathrm{1:1\ E}}^{t_i}$ has a weight of $(\Delta t_i/\Delta T)$ for average.

\subsection{Source crater criterion.}
For any possible source crater with diameter and age, we estimate its corresponding impactor size according to Eq.~(\ref{eq_scaling1}) and Eq.~(\ref{eq_scaling2}), considering the same impact velocity as our simulations.
The impactor mass is then related to the total escaping fragments, assuming a linear dependence as Eq.~(\ref{eq_escape_imp}).
Since the Earth co-orbital fraction does not follow a simple power-law relationship over time (Fig.~3b), we fit it with a cubic function
\begin{equation}\label{eq_R2L}
    \ln(R_{\mathrm{1:1\ E\ to\ Launch}}(t)) = \sum_{i=0}^{3} \alpha_i \left[\ln(t)\right]^i
\end{equation}
where the time $t$ is in Myr, the coefficients are given in Supplementary Table~5.

Combining the above relations allows us to estimate the Earth co-orbitals number for any possible source crater with specific diameter and age,
\begin{equation}\label{eq_1E}
    N_{\mathrm{1:1\ E}} = N_{\mathrm{Launch}} \cdot R_{\mathrm{1:1\ E\ to\ Launch}}
\end{equation}
To obtain at least one Kamo`oalewa-sized Earth co-orbital asteroid at the present day, the source crater's size $D_{\mathrm{final}}$ and age $t$ must satisfy
\begin{equation}\label{eq_1E_}
    N_{\mathrm{1:1\ E}} \geq 1
\end{equation}
and the results are shown in Fig.~4.

\end{methods}

\begin{addendum}
\item[Data availability]
The data that support the plots within this paper have been provided as Source Data. The raw simulation data are available from the corresponding authors upon reasonable request.

\item[Code availability]
The spectral classification code classy is open access at \href{https://github.com/maxmahlke/classy}{https://github.com/maxmahlke/classy}.
The N-body code REBOUND is an open-source package at \href{https://github.com/hannorein/rebound}{https://github.com/hannorein/rebound}.
The impact code SPHSOL is available from the corresponding authors on reasonable request.


\item[Acknowledgements]
B.C. is supported by the National Natural Science Foundation of China (No. 12202227) and the Postdoctoral Innovative Talent Support Program of China (No. BX20220164). This work is also supported by the National Natural Science Foundation of China under Grant 62227901. We thank Dr. William F. Bottke and others for valuable discussions on this work at the Asteroids, Comets, Meteors Conference (ACM 2023). We thank M. Connors, T. Santana-Ros and F. Ferrari for providing helpful comments to improve and clarify the manuscript. We acknowledge the use of imagery from Lunar QuickMap (\href{https://quickmap.lroc.asu.edu}{https://quickmap.lroc.asu.edu}), a collaboration between NASA, Arizona State University \& Applied Coherent Technology Corp.

\item[Author contributions]
Y.J. performed the SPH and N-body numerical simulations and analysed the numerical results. B.C. and H.B. initiated the project, designed the simulations and led the research. Y.H., B.G. and R.M. contributed to the discussion of the dynamical evolution of lunar ejecta and the spectral comparison. E.A., P.M. and Y.Y. contributed to the discussion of the lunar impact ejection process. All authors contributed to interpretation of the results and preparation of the manuscript.

\item[Competing Interests]
The authors declare no competing interests.

\end{addendum}

\clearpage
\begin{figure}[!htb]
    \includegraphics[width=0.6\textwidth]{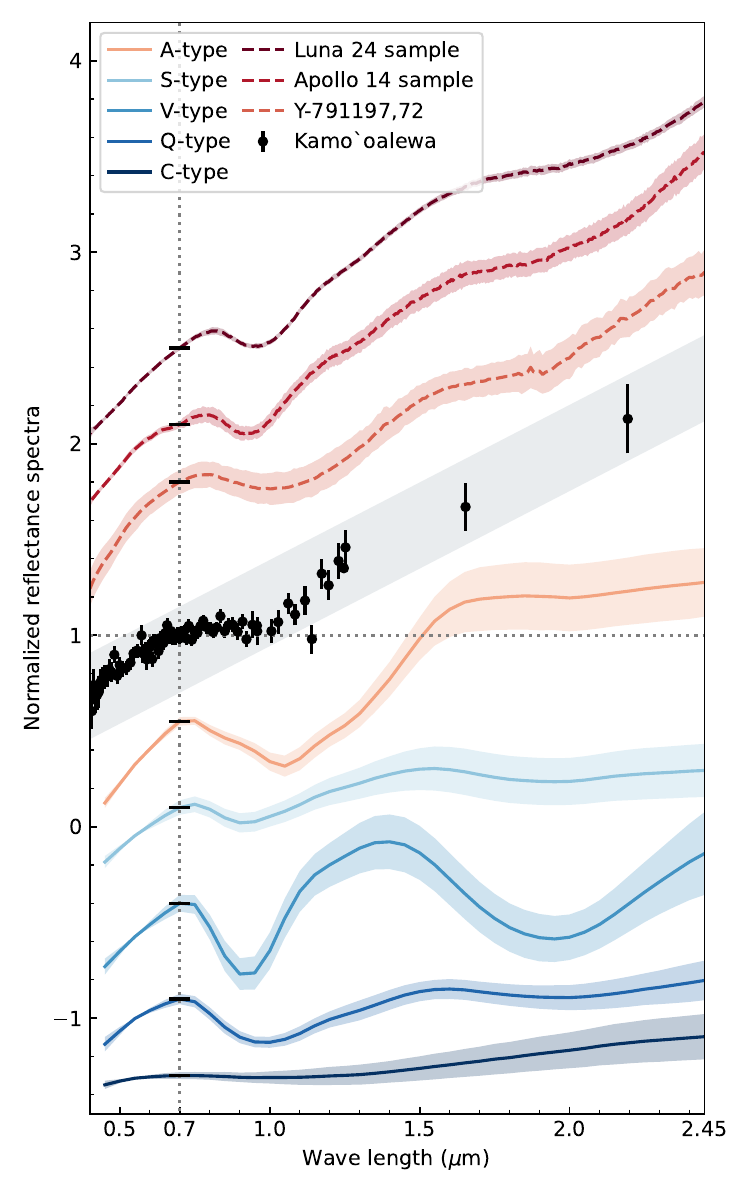}
    \centering
    \caption{\textbf{Kamo`oalewa's observed reflectance spectrum\cite{sharkey2021lunar} compared with typical asteroid taxonomies\cite{demeo2009extension}, lunar samples\cite{relab} and {the lunar meteorite Yamato-791197,72}\cite{relab}.}
    For Kamo`oalewa, the error bars represent the photometric uncertainties and the zJHK color ratio measurements\cite{sharkey2021lunar}.
    For other spectra, the mean spectrum (solid lines for asteroid types and dashed lines for lunar materials) and standard deviation (shaded area) are plotted, where the statistics are based on 371 asteroid samples\cite{demeo2009extension} and the specified lunar materials measurements\cite{relab}.
    All spectra have been firstly normalized to unity at 0.7 $\mu$m, and then shifted vertically for clarity, {as indicated by the short horizontal lines}. The grey area indicates Kamo`oalewa's mean slope of 89\%$/\mu$m\cite{sharkey2021lunar}. See Extended Data Fig.~1 for slopes comparison\cite{demeo2009extension}, Extended Data Fig.~2 for classifications using the {Mixture of Common Factor Analysers (MCFA)} model\cite{mahlke2022asteroid}.
    }
    \label{fig1}
\end{figure}

\clearpage
\begin{figure}[!htb]
    \includegraphics[width=1.0\textwidth]{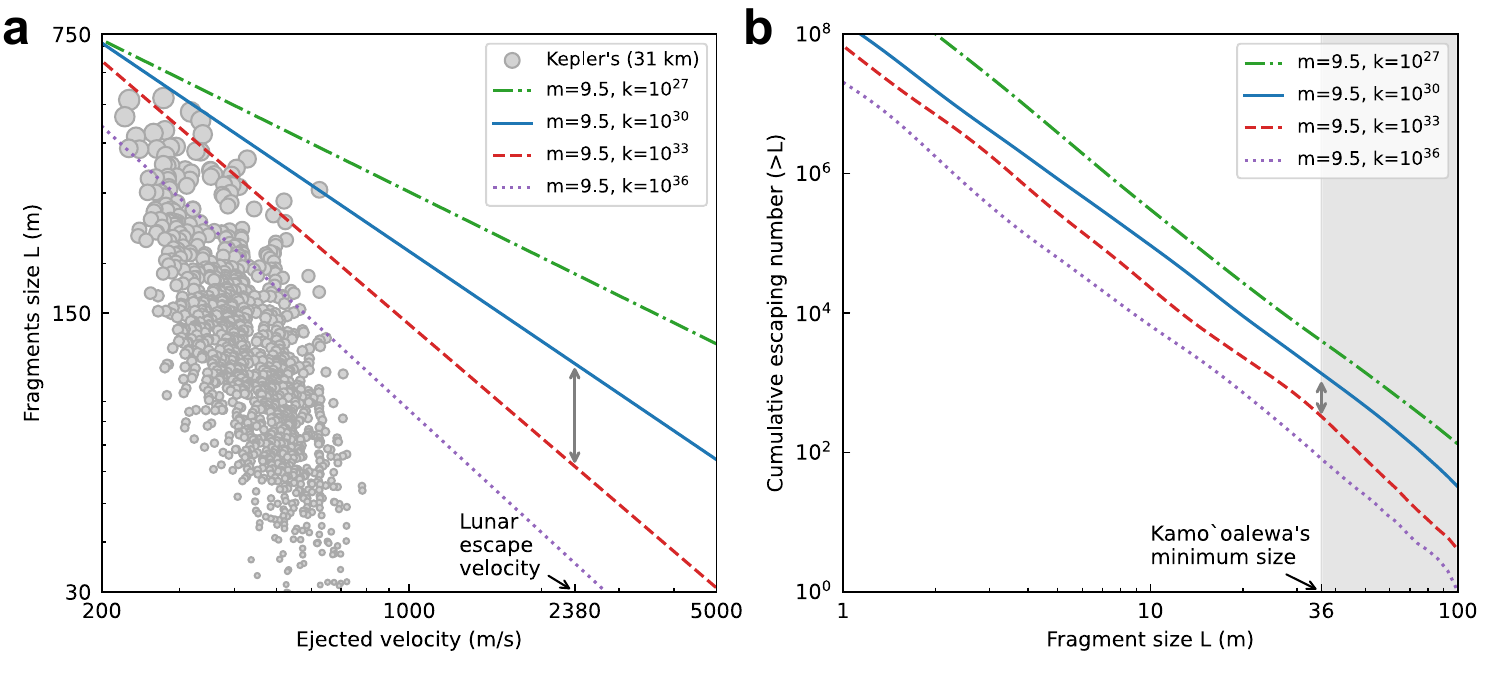}
    \centering
    \caption{\textbf{Simulated {fragment size-velocity distribution (SVD) and escaping fragment cumulative size-frequency distribution (CSFD)} for a 2.44-km-diameter asteroid striking the lunar crust at 18 km/s and 45$^\circ$, creating a Kepler-sized crater (31 km).}
    \textbf{a,} {The lines represent the largest fragment sizes versus ejected velocities with different Weibull parameters, fitted as a simple power law.}
    The circle points are computed from the secondary craters of the Kepler crater\cite{singer2020lunar}, which has a cutoff at about 800 m/s due to the difficulty in searching secondaries at great distances. 
    The upper limit of the Kepler's fragment SVD helps us to estimate the Weibull parameter $k$ ranging from 10$^{30}$ to 10$^{33}$ m$^{-3}$, and thereby the largest escaping fragment size ranging from 62 to 112 m, or 2.5--4.6\% of the impactor size (the double arrow).
    \textbf{b,} CSFD of ejected fragments with velocities faster than lunar escape velocity. As illustrated by the arrow, we estimate about 1250 escaping fragments ($>$36 m) for $k$ = 10$^{30}$ m$^{-3}$, and 300 for $k$ = 10$^{33}$ m$^{-3}$, with $m$ = 9.5. 
    The fitted slopes range from $-$3.4 to $-$3.6, which is close to the ejected boulders distribution around lunar craters\cite{bart2010distributions}, so the debris is dominated by smaller escaping fragments.
    }
    \label{fig2}
\end{figure}

\clearpage
\begin{figure}[!htb]
    \includegraphics[width=1.0\textwidth]{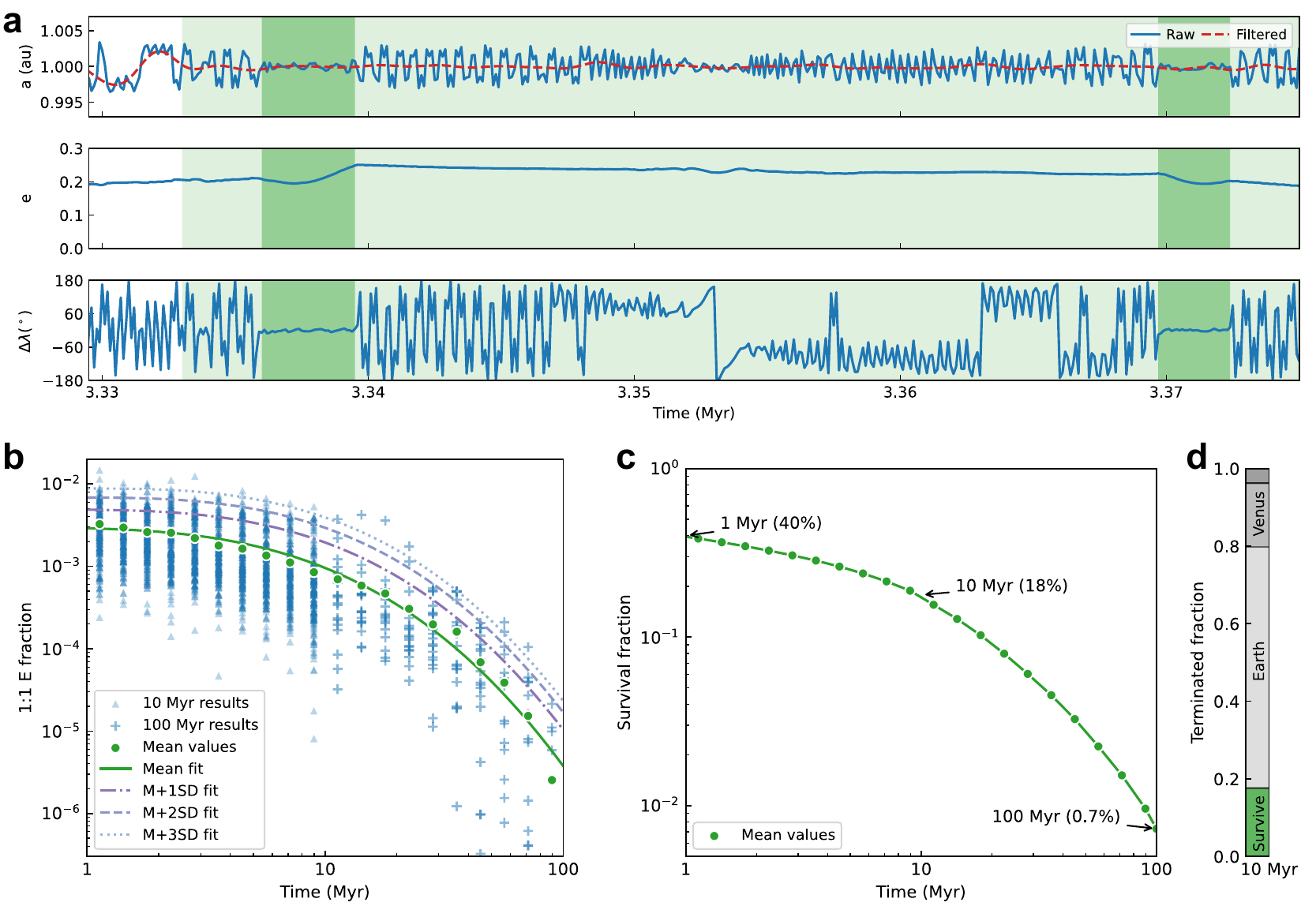}
    \centering
    \caption{\textbf{N-body simulation results of Giordano Bruno (GB) ejecta.}
    \textbf{a,} Example of Earth co-orbital states in dynamical evolution. The green shaded area represents QS state like Kamo`oalewa' recent orbit. And the light green shaded area represents the other two types of Earth co-orbital configurations: HS and {TP (Tadpole, also known as Earth Trojans\cite{connors2011earth,santana2022orbital})}. See Supplementary Fig.~8 for its orbital change over total 10 Myr.
    \textbf{b,} Earth co-orbital fraction over time in a log-log plot. As defined by Eq.~(\ref{eq_define_R2L}) in Methods, we use the average value over a time interval $\Delta T$ to evaluate the proportion of the Earth co-orbitals to the launched particles. The time intervals are chosen to give a constant bin size on a logarithmic scale. The triangular points represent the results of 10-Myr simulations, and the pluses represent the additional 100-Myr simulations. 
    Since the Earth co-orbital fraction does not follow a simple power law over time, we suppose a cubic relation to fit the mean values (solid line) and the standard deviations.
    \textbf{c,} The survival fraction of GB ejecta in total 100 Myr, calculated as the mean values of all simulated GB events.
    {\textbf{d,} Fates of GB ejected particles at 10 Myr. Only 18\% of all launched particles survive to 10 Myr. Most of the particles impacted Earth ($\sim$60\%), and about 16\% hit Venus. See Supplementary Table~3 for more information.}}
    \label{fig3}
\end{figure}

\clearpage
\begin{figure}[!htb]
    \includegraphics[width=0.75\textwidth]{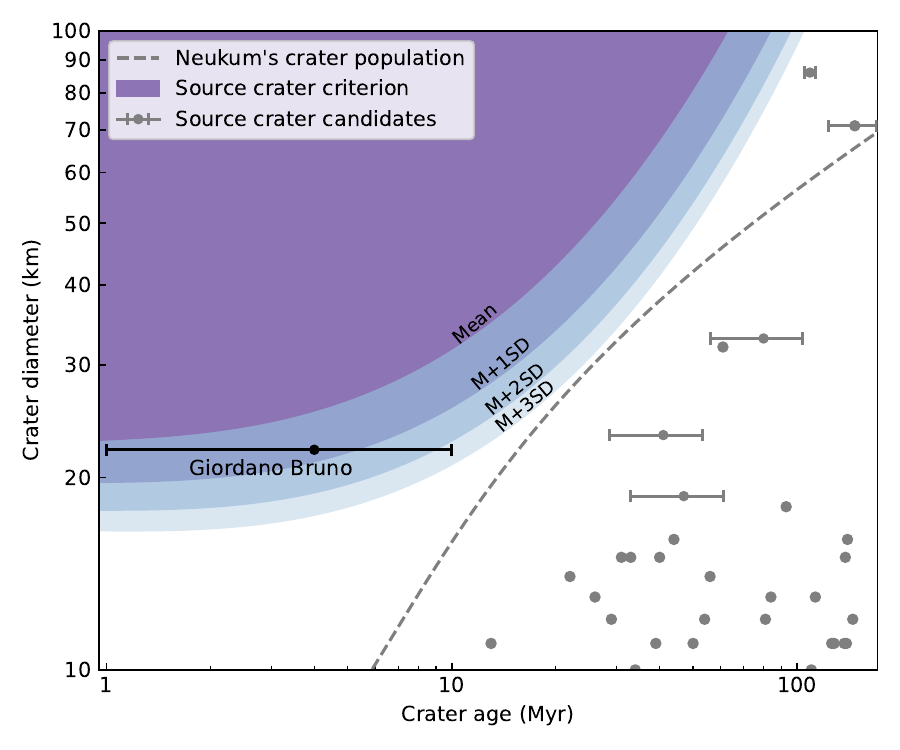}
    \centering
    \caption{\textbf{Source crater criterion to obtain at least one Earth co-orbit at the present day.}
    Using our simulated results as a baseline, we can further constrain all possible source craters based on their estimated ages and sizes.
    The areas labeled as ``Mean'', ``M+1SD'', ``M+2SD'' and ``M+3SD'' are calculated from the fitted curves in Fig.~3b and represent the uncertainty in Earth co-orbital results due to the chaotic dynamics, based on our 220 sets of N-body simulations.
    The dashed line shows the nominal largest crater size over time according to Neukum's model\cite{neukum2001cratering}, indicating that GB is unusually larger than the average. The other source crater candidates are listed in Supplementary Table~2. The error bars represent the mean values and uncertainties in the carter age, based on either small crater size-frequency measurements\cite{morota2009formation} or rock cosmic-ray exposures\cite{drozd1977cosmic}.
    }
    \label{fig4}
\end{figure}

\clearpage

\end{document}